\documentclass[aps,prb,twocolumn,superscriptaddress,showpacs,floatfix]{revtex4}
\usepackage{graphicx}
\usepackage{amsmath}
\usepackage{amssymb}
\usepackage{url}

\newcommand{\E}{\ensuremath{{\mathrm{e}}}}
\newcommand{\imag}{\ensuremath{{\mathrm{i}}}}

\graphicspath{{.}{../}{./eps/}}

\begin{document}

\title{Anderson disorder in graphene nanoribbons: A local distribution approach}
\author{Gerald Schubert}
\affiliation{Institut f\"ur Physik, Ernst-Moritz-Arndt Universit\"at
  Greifswald, 17487 Greifswald, Germany}
\affiliation{Regionales Rechenzentrum Erlangen, 91085 Erlangen, Germany}
\author{Jens Schleede}
\author{Holger Fehske}
\affiliation{Institut f\"ur Physik, Ernst-Moritz-Arndt Universit\"at
  Greifswald, 17487 Greifswald, Germany}
\date{\today}
\begin{abstract}
  Disorder effects strongly influence the transport properties of 
  graphene based nanodevices even to the point of Anderson localization. 
  Focusing on the local density of states and its distribution function, 
  we analyze the localization properties of actual size graphene nanoribbons.
  In particular we determine the time evolution and localization length of 
  the single particle wave function in dependence on the ribbon extension 
  and edge geometry, as well as on the disorder type and strength.
\end{abstract}

%  72.15.Rn Localization effects (Anderson or weak localization)
%  71.23.An Theories and models; localized states
%  05.60.Gg Quantum transport
%  71.30.+h Metal-insulator transitions and other electronic transitions

\maketitle

\section{Introduction}

Disorder effects in graphene are of particular importance on the account of 
its two-dimensional (2D) lattice structure. 
The single parameter scaling theory predicts that in 2D systems  
arbitrary weak disorder leads to Anderson localization of the single
particle wave function~\cite{LR85}. 
For graphene it has been argued that due to the linear dispersion in 
the vicinity of the band center the one parameter scaling theory does 
not hold. The problem of Anderson localization in graphene 
is therefore heavily debated~\cite{OGM06,NKR07,BTBB07}.

Accessing Anderson localization both theoretically and experimentally, the 
local density of states (LDOS) is a central quantity.
By means of the local distribution approach, the distribution of the
LDOS may be used to distinguish localized from extended states~\cite{AF08c,STBB02,Mi00}.
Nowadays, the LDOS can be directly measured by scanning 
tunneling spectroscopy experiments~\cite{NKF09,NKMYF06,MKNTTF05,MKMW03}.

An ordered, infinite graphene sheet is a zero-gap semiconductor with 
a linear density of states near the charge neutrality point~\cite{Wa47}.
Cutting a graphene nanoribbon (GNR) of finite width out of such a sheet, 
additional aspects  have to be considered. 
First, the finite number of transverse atoms causes quantum confinement,
% and opens a gap in the band center.
%
where the presence of the edges leads to a symmetry breaking. 
%introduces disorder in the system. 
Second, lattice defects or targeted implementations of e.g. boron ($\rm B_7$) 
clusters~\cite{QOKF08} result in random samples.    
Thereby the range of the disorder is of great 
importance~\cite{CNBNTCR08}:
For long-range disorder, as caused by ripples in the graphene sheet, 
the two independent corners of the Brillouin zone are untangled and 
long-wavelength excitations can be modeled by an effective Dirac 
equation.
If the scattering potential is short ranged, however, 
inter-valley scattering between the two inequivalent
Dirac cones becomes possible.
Third, the finite extension (aspect ratio) of the GNR introduces  
a new length scale being absent in infinite graphene sheets.
Actually we expect metallic behavior of disordered quasi-1D GNRs 
if the localization length becomes comparable or even larger than the 
longitudinal ribbon size~\cite{AGW07,LBNR08,NRC08}.

To address these questions, in this work we investigate 
the electronic structure and the localization properties of 
disordered GNRs by means of unbiased numerical techniques.
Thereby we focus on the interplay of %substitutional 
disorder, boundaries effects and GNR geometry.
Particular aspects of various kinds of disorder in GNRs have been 
investigated previously in the 
literature ~\cite{AGW07,LBNR08,NRC08,QAVHBGD08,LL08,EZXH08,MCL09,TUHSG09p}.

\section{Model and methods}

To this end we consider the tight-binding Hamiltonian
\begin{equation}\label{H_bm}
  {H} =  \sum_{i=1}^{N} \epsilon_i {c}_i^{\dag} {c}_i^{} 
  - \bar{t} \sum_{\langle ij \rangle}({c}_i^{\dag} {c}_j^{} + \text{H.c.})
\end{equation}
on a honeycomb lattice with $N$ sites, including hopping between
nearest neighbors $\langle ij\rangle$ only. 
Drawing the on-site potentials $\epsilon_i$ from the box distribution 
\begin{equation}
 p[\epsilon_i] = 
\frac{1}{\gamma}\;\theta\left(\gamma/2-|\epsilon_i|\right)\,,
\end{equation}
we introduce (short ranged) Anderson disorder~\cite{An58}.
We distinguish between bulk ($\gamma_b$) and edge ($\gamma_e$) disorder, 
when all on-site potentials are subjected to $p[\epsilon_i]$
or only those at the edge sites. 
We consider quasi-1D GNRs of finite widths with open (periodic) 
boundary conditions in transverse (longitudinal) direction.
Depending on the orientation of the GNRs with respect to the honeycomb
lattice, zigzag or armchair geometries will be realized 
(see panels on top of Fig.~\ref{fig:GNR_mety}).

The local properties of site $i$ of a %disordered 
sample with broken translational invariance 
are reflected in the LDOS, 
\begin{equation} \label{LDOS}
  \rho_i(E) = \sum\limits_{m=1}^{N}
  | \langle i | m \rangle |^2\, \delta(E-E_m)\;.
\end{equation}
Recording the probability density function $f[\rho_i]$ for many 
different sites $\{i\}$ of a given sample and different sample 
realizations $\{\epsilon_i\}$ restores translational invariance
on the level of distributions:
The shape of $f[\rho_i]$ is determined by $p[\epsilon_i]$ but
independent of $\{i\}$ and $\{\epsilon_i\}$~\cite{AF08c}.
For extended states, $f[\rho_i]$ is strongly peaked around 
the mean DOS, 
\begin{equation}
   \rho_{\text{me}}=\langle \rho_i\rangle\;, 
\end{equation}
independent 
of the system size, whereas  for localized states $f[\rho_i]$ exhibits a
log-normal distribution that becomes singular for increasing system 
sizes~\cite{swwf05}.
Normalizing the LDOS to $\rho_{\text{me}}$ 
allows for a detection of the localization properties by performing a
finite size scaling for the LDOS distribution.
More conveniently, the typical DOS 
\begin{equation}
  \rho_{\text{ty}}=\E^{\langle \ln \rho_i\rangle}
\end{equation}
monitors the changes in the LDOS distribution.
While for $N\to\infty$ an extended state is characterized by finite values
of $\rho_{\text{me}}$ and $\rho_{\text{ty}}$, for localized states 
$\rho_{\text{me}}$ is finite but $\rho_{\text{ty}}\to 0$~\cite{swwf05}.

Alternatively, the recurrence probability $P_R(t\to\infty)$
also reveals the localization properties of the system~\cite{KM93_2}.
While in the thermodynamic limit $P_R\sim1/N\to0$ for extended states,
localized states are characterized by a finite value of $P_R$. 
Starting from a localized wave packet, we are able to calculate the 
time dependent local particle density,
\begin{equation}
  n_i(t) = |\psi({\mathbf r}_i, t)|^2 = \Big| \sum\limits_{m=1}^{N}  
  \E^{-\imag E_m t} \langle m| \psi(0)\rangle \langle i| m \rangle \Big|^2\;,
\end{equation}
by expanding the time evolution operator into a finite series of Chebyshev
polynomials~\cite{TK84,WF08c}.
The above local distribution approach also applies to $n_i(t)$.
But since an initial state in general contains contributions of the whole
spectrum, examining $n_i(t)$ does not allow for an energy resolved 
investigation of localization as by the LDOS.  
Instead it provides a tool for a global examination of the spectrum with
relevance for possible measurements.
Note that a finite overlap of just one extended state with the initial state
leads to a complete spreading of this state after some time.

\section{Numerical results}
\subsection{Local density of states}

\begin{figure}%[htbp]
  \hspace*{0.08\linewidth}\includegraphics[width=0.8\linewidth,clip]{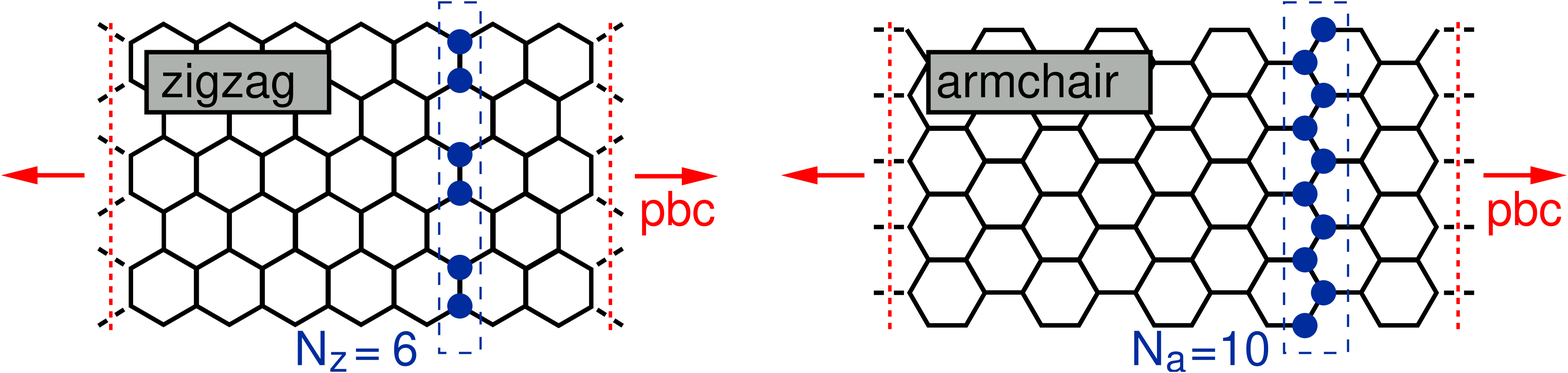}
  \centering\includegraphics[width=0.96\linewidth,clip]{fig1b.eps}
  \caption{(Color online)
    Mean (solid red) and typical (dashed blue) DOS for zigzag (left column, $N_z=6$) and 
    armchair (right column, $N_a=10$) GNRs of width $W=1.1\,\mathrm{nm}$. 
    Top panels: Ordered case.
    Lower panels: In each $(2\times2)$ block,
    we compare for a fixed value of disorder the influence of 
    bulk disorder ($\gamma_b$, upper rows) to edge disorder
    ($\gamma_e$, bottom rows). 
    To illustrate the localization properties, in each panel 
    $\rho_{\text{ty}}$ is given for $L = 213\,(1064)\,\mathrm{nm}$ by 
    lightblue dot-dashed (darkblue dashed) lines.
    These system sizes correspond to $10000$ $(50000)$ lattice 
    sites for the armchair and $10392$ $(51960)$ for the zigzag case.
    Disorder averaging was performed over $10^5$ realizations.
    In the longitudinal direction periodic boundary conditions (pbc) are
    applied.}
  \label{fig:GNR_mety}
\end{figure}

Compared to the band structure of an infinite 2D %two dimensional 
graphene sheet, the DOS of finite GNRs is characterized by a 
multitude of van Hove singularities (see top panels of 
Fig.~\ref{fig:GNR_mety}).
For zigzag GNRs, the strong signature at $E=0$ indicates the high 
degeneracy of the edge states~\cite{NFDD96}.
In contrast, armchair GNRs with $N_a=3n$ or $N_a=3n+1$ are gapped
around $E=0$. % for arbitrary integers $n$. 
This finite size gap tends to zero as $N_a\to\infty$.
The resulting metallicity for $N_a=3n+2$ is an artifact of the NN
tight binding approximation, however, and vanishes if
next- and third NN are taken into account~\cite{SCL06c}.
For other values of $N_a$ a longer ranged hopping slightly modifies 
the gap size but does not change the fundamental behavior.
Note that even for vanishing Anderson disorder the LDOS varies for 
different bulk sites according to their relative position  
to the ribbon edges.
Symmetry considerations show that there are $N_z$ ($N_a/2$) inequivalent
lattice sites in ordered zigzag and armchair GNRs.
Therefore mean and typical DOS do not coincide even for $\gamma_b=0$
(see, e.g., the band center of the zigzag GNR). 

If disorder comes in, localized states emerge in the band gap of the
armchair GNRs, and above a critical disorder strength the gap
is filled completely. 
The localization properties of the states can readily be seen from
the system size dependence of $\rho_{\text{ty}}$.
A tendency towards reduced values of $\rho_{\text{ty}}$ for increasing
system sizes indicates localization for both GNR geometries 
and all energies. %the shown values of disorder.
While this localization effect arises for bulk disorder already at 
$\gamma_b/\bar{t}=2$, an edge disorder strength of $\gamma_e/\bar{t}=2$ is
still too weak to localize the wave function on GNRs of 
$L = 213\,\mathrm{nm}$ size as indicated by the approximate equality of 
$\rho_{\text{ty}}$ and $\rho_{\text{me}}$.
A substantial reduction of $\rho_{\text{ty}}$ is only observed for 
larger systems ($L = 1064\,\mathrm{nm}$) which indicates localization
on a larger length scale.
Obviously, zigzag GNRs are less sensible to edge disorder than 
armchair GNRs since this geometry has only half the number
of (disordered) edge sites.
The different edge geometries are only of importance if the disorder is weak. 
For strong disorder, $\gamma_b/\bar{t}=4$, the results for armchair and zigzag 
GNRs coincide almost exactly.

\begin{figure}%[htbp]
  \centering\includegraphics[width=\linewidth,clip]{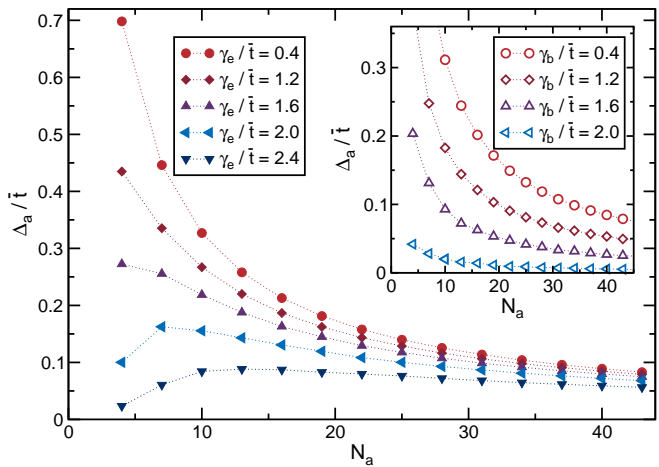}
  \caption{(Color online) Gap size $\Delta_a$  for armchair 
    GNRs as a function of ribbon width $N_a=3n+1$ for edge disorder 
    (main panel) and bulk disorder (inset). 
   Results are based on the averaged DOS for GNRs with a length 
    of $1000$ atoms using $4096$ realizations of disorder.} 
  \label{fig:gapsize_W_Na}
\end{figure}

As stressed above, there are three branches of gap sizes depending on
$\text{mod}(N_a,3)$. 
In Fig.~\ref{fig:gapsize_W_Na} we focus on $N_a=3n+1$ and examine the 
influence of both bulk and edge disorder on the gap size $\Delta_a$
in dependence on the ribbon width.
For our finite system we calculate $\Delta_a$ as 
\begin{equation}
  \int\limits_{-\Delta_a/2}^{\Delta_a/2} \rho_{\text{me}}(E)\, dE = \frac{1}{N}\;.
\end{equation}
A finite--size analysis shows that upon increasing the ribbon width the gap 
narrows for any bulk disorder $\gamma_b$.
In contrast, for edge disorder we observe a non-monotonic behavior that
can be explained by the competition of two effects:
Increasing the width of the GNR on the one hand weakens the influence of the 
disorder as the ratio of edge to bulk sites decreases. 
An increasing number of lattice sites, on the other hand, 
reduces the finite size effects and closes the gap.
Thus, for $\gamma_e\gtrsim 2\bar{t}$, the gap first broadens when 
the GNR width is increased, and then converges to the gap size of the 
ordered system, which finally vanishes in the limit $N_a\to\infty$.
Similar studies for a different type of edge disorder, in which 
sites are randomly removed from the ribbon edges, can be found in the 
literature~\cite{MCL09,EZXH08,QAVHBGD08}.  
\begin{figure}%[htbp]
  \centering\includegraphics[width=\linewidth,clip]{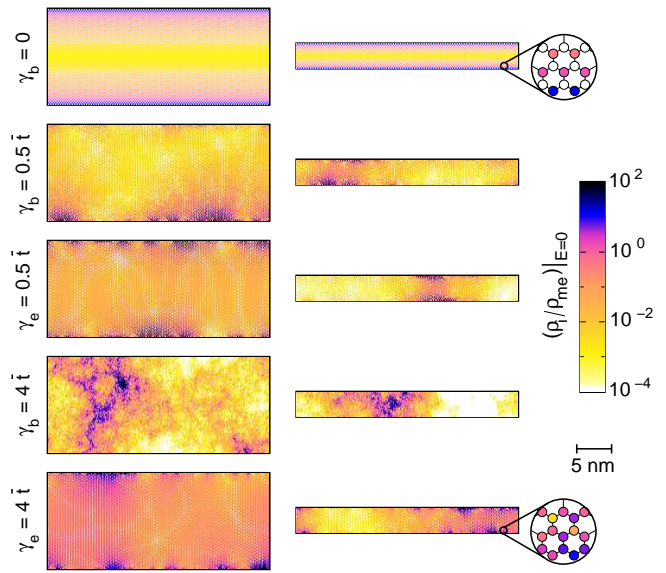}
  \caption{(Color online) Normalized LDOS at the band center
    $(\rho_i/\rho_{\text{me}})|_{E=0}$
    for particular zigzag GNRs. In addition to contrasting bulk and edge 
    disorder, we compare in the left(right) column the influence of the 
    aspect ratios $L \times W= 31.4\times 13.5 \, (31.4 \times 3.3) \, 
    \text{nm}^2$, corresponding to $256 \times 64 \, (256 \times 16)$ sites. 
    Results obtained by exact diagonalization.}
  \label{fig:GNR_CharStates}
\end{figure}

To get further insight into the nature of the eigenstates of GNRs 
and substantiate our conclusions about their localization properties, 
we show the LDOS in the band center in Fig.~\ref{fig:GNR_CharStates}.
The magnifying inset for the ordered case shows the alternating structure 
of the edge states which are distinctive for the band center of zigzag 
GNRs~\cite{NFDD96}.  
In the presence of weak edge disorder, the checkerboard structure of the
amplitudes persists in the bulk, while near the edges regions with 
significantly enhanced amplitudes emerge.
The A-B sublattice structure is no longer present
for larger $\gamma_e$ as can be seen in the lower inset of 
Fig.~\ref{fig:GNR_CharStates}.
Here, the sites with vanishing amplitudes form a filamentary network in the 
bulk, caused by the influence of the disordered edges.
For bulk disorder, localization arises first near the edges of the system
in the case of weak disorder, while localized states in the bulk of the 
GNR occur only for strong disorder. 
Varying the aspect ratio of the GNRs (right column of 
Fig.~\ref{fig:GNR_CharStates}), we may tune the relative importance of
the edges in the system. 
Although this effect is most pronounced for edge disorder, we observe also for 
bulk disorder such a  ``renormalization'' of the disorder
strength: A given $\gamma_{b,e}$ causes stronger localization for 
narrow GNRs.

\subsection{Time evolution of the wavefunction}
\begin{figure}%[htbp]
  \centering\includegraphics[width=\linewidth,clip]{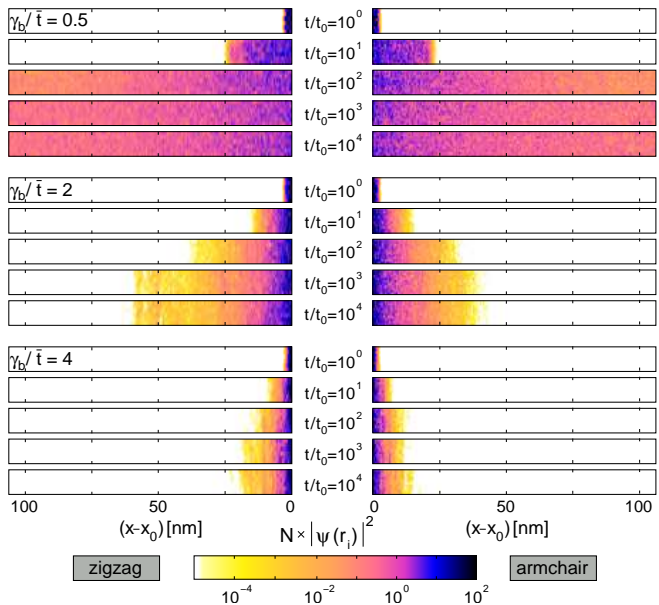}
  \caption{(Color online) Time evolution of the normalized particle density 
    $N |\psi({\mathbf r}_i)|^2$ on disordered GRNs with 
    zigzag and armchair geometries for different values of bulk disorder 
    $\gamma_b$. Device dimensions: $(1.1\times213)\,\text{nm}^2$ corresponding 
    to $6\times 1732$ atoms (zigzag) and $10\times 1000$ atoms (armchair).
    Times are measured in units of the inverse hopping element 
    $t_0=1/\bar{t}$.}
  \label{fig:GNR_TimeEv}
\end{figure}

Figure~\ref{fig:GNR_TimeEv} shows the time evolution of an initially 
localized state, as calculated by the Chebyshev method~\cite{WF08c}.
The dynamics of the initial wave packet is characterized by a fast 
spreading process ($t\lesssim10^3t_0$), after which its extension 
does not change anymore, even for very long times.
Clearly, on individual sites the amplitudes fluctuate in time, 
but the overall nature of the state for $t=10^4t_0$ is quasistationary.
The localization properties depend on both disorder strength and edge 
geometry. 
Obviously, armchair GNRs are more susceptible to the presence of 
disorder than those of zigzag type. % for the same $\gamma_b$.
For the shown GNRs of moderate length and weak disorder 
($\gamma_{b,e}/\bar{t}=0.5$) the localization length is larger than the
system size and thus the GNR is ``metallic''.

\begin{figure}%[htbp]
  \centering\includegraphics[width=\linewidth,clip]{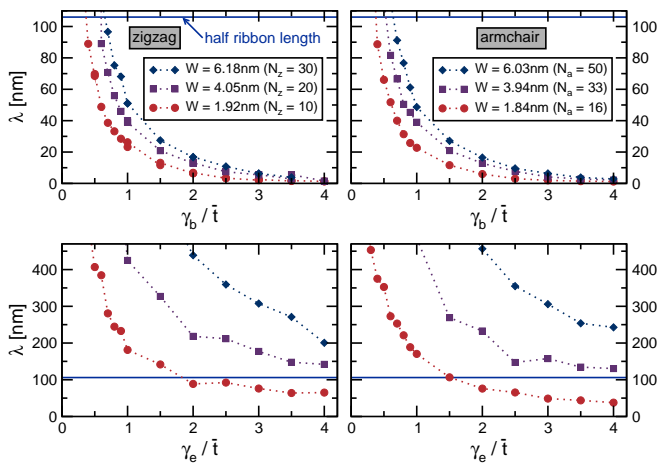}
  \caption{(Color online) 
    Localization length in dependence on bulk ($\gamma_b$) and edge ($\gamma_e$)
    disorder strength for armchair and zigzag GNRs.
    The values are sample averages obtained for 10 GNRs of 
    $L=213\,\text{nm}$ when the state has become quasistationary. 
  }
  \label{fig:GNR_LL}
\end{figure}

The extraction and quantitative discussion of the localization
length in narrow GNRs is challenging.
There is no problem to determine $\lambda$ from an exponential fit 
\begin{equation}
|\psi(\mathbf{r}_i)|^2 = |\psi(\mathbf{r}_0)|^2
\exp\left(-\frac{|\mathbf{r}_i-\mathbf{r}_0|}{\lambda}\right) 
\end{equation}
for
a given initial state and disorder realization at any fixed time.
But the such-determined $\lambda$ strongly fluctuates, both in time
and as a function of the chosen initial state and disorder realization.
The temporal fluctuations of about 5-10\% can be eliminated by 
time averaging.
Varying the initial state and/or comparing different disorder 
realizations, leads to additional uncertainties of about 10-20\%.
Therefore we show in Fig.~\ref{fig:GNR_LL} sample averages over 
several combinations of initial states and disorder realizations.

Figure~\ref{fig:GNR_LL} indicates that the influence of the 
boundary (armchair/zigzag) is only of minor importance for the 
localization length.
But we observe a pronounced difference between bulk and 
edge disorder, with $\lambda>L$ also for large values of $\gamma_e$
for most ribbon widths. 
For any fixed disorder strength, a decreasing width of the GNR systematically 
reduces $\lambda$ since the influence of the lateral dimension is
weakened and the system approaches the 1D limit.
Values of $\lambda$ which are significantly larger than half the system
size (blue solid line) have to be taken with care since a reliable
determination of the localization length requires $\lambda\lesssim L$.
Clearly, the precise value of $\lambda$ in those cases is of minor 
importance due to the metallic behavior of such finite GNRs.
A quantitative comparison of the obtained localisation lengths 
with estimates based on other methods~\cite{LBNR08,AGW07,NRC08}
suffers from the different investigated disorder models.
Nevertheless, the orders of magnitude match and the general tendencies
are reproduced:
the impact of disorder increases with decreasing ribbon width and
the boundary type does not influence the localization length 
significantly for strong disorder.
The pronounced dependence of the localization length on the ribbon type 
(armchair or zigzag) for the weakly disordered case reported 
in Refs.~\onlinecite{LBNR08,AGW07} is absent in our data.
We attribute this to the different disorder models used.

\section{Summary}

To conclude, Anderson localization takes place in disordered quasi-1D 
graphene nanoribbons, but taking into account the actual device dimensions 
GNRs can be conducting at weak disorder strengths. 
This has been proven by calculating the localization length
and time evolution of single particle states. 
Within the local distribution approach Anderson localization 
is identified by a log-normal distribution of the LDOS 
that shifts towards zero for increasing system size. 
The LDOS is directly measurable by scanning tunneling spectroscopy  
and therefore allows for a direct comparison of theory and experiment.

\section*{Acknowledgments}

This work was funded by the Deutsche Forschungsgemeinschaft through 
the Research Program SFB TR 24 and the Competence Network for 
Technical/Scientific High-Performance Computing in Bavaria (KONWIHR).
The numerical calculations have been performed on the Tera\-Flop compute 
cluster at the Institute of Physics, Greifs\-wald University.

\vspace*{-1cm}
%\bibliography{./ref} 

\bibliographystyle{apsrev}

\end{document}